\documentclass[twoside,epsf]{article}

\usepackage{amsmath}
\usepackage{lscape}
\usepackage{longtable}

\newcommand\blfootnote[1]{%
  \begingroup
  \renewcommand\thefootnote{}\footnote{#1}%
  \addtocounter{footnote}{-1}%
  \endgroup
}

\newcommand*\aap{A\&A}

\newcommand*\aaps{A\&AS}

\newcommand*\aj{AJ}
\newcommand*\apj{ApJ}
\newcommand*\apjl{ApJ}

\newcommand*\apjs{ApJS}

\newcommand*\memsai{Mem.~Soc.~Astron.~Italiana}
\newcommand*\mnras{MNRAS}
\newcommand*\na{New A}

\newcommand*\pasp{PASP}

\input ibvs2.sty

\begin{document}

\IBVShead{6xxx}{00 Month 2016}

\IBVStitletl{White Dwarf Period Tables}{I. Pulsators with hydrogen-dominated atmospheres}

\IBVSauth{Zs. Bogn\'ar; \'A. S\'odor}

\IBVSinsto{Konkoly Observatory, MTA Research Centre for Astronomy and Earth Sciences, Konkoly Thege Mikl\'os \'ut 15-17, H--1121 Budapest}
\IBVSinsto{e-mail: bognar.zsofia@csfk.mta.hu}

\IBVStyp{ ZZ }
\IBVSkey{Catalogs, Stars: oscillations, white dwarfs}

\SIMBADobj{SDSS J000006.75-004654.0}
\SIMBADobj{SDSS J001836.11+003151.1}
\SIMBADobj{WD 0016-258}
\SIMBADobj{HE 0031-5525}
\SIMBADobj{WD 0036+312}
\SIMBADobj{WD 0041+006}
\SIMBADobj{SDSS J004855.17+152148.7}
\SIMBADobj{WD 0049-473}
\SIMBADobj{SDSS J010207.17-003259.4}
\SIMBADobj{WD 0104-464}
\SIMBADobj{SDSS J011100.63+001807.2}
\SIMBADobj{WD 0120+002}
\SIMBADobj{SDSS J012950.44-101842.0}
\SIMBADobj{WD 0132-014}
\SIMBADobj{WD 0133-116}
\SIMBADobj{WD 0145-221}
\SIMBADobj{SDSS J021406.78-082318.4}
\SIMBADobj{WD 0235+069}
\SIMBADobj{SDSS J024922.35-010006.7}
\SIMBADobj{WD 0246+326}
\SIMBADobj{SDSS J030153.81+054020.0}
\SIMBADobj{SDSS J030325.22-080834.9}
\SIMBADobj{SDSS J031847.09+003029.9}
\SIMBADobj{SDSS J033236.61-004918.3}
\SIMBADobj{WD 0341-459}
\SIMBADobj{WD 0344+073}
\SIMBADobj{HE 0344-1207}
\SIMBADobj{SDSS J034939.35+103649.9}
\SIMBADobj{WD 0416+272}
\SIMBADobj{WD 0417+361}
\SIMBADobj{WD 0455+553}
\SIMBADobj{LP 119-10}
\SIMBADobj{WD 0507+045}
\SIMBADobj{WD 0517+307}
\SIMBADobj{WD 0532-560}
\SIMBADobj{PM J07029+4406}
\SIMBADobj{SDSS J073707.99+411227.6}
\SIMBADobj{SDSS J075617.54+202010.2}
\SIMBADobj{SDSS J081531.75+443710.3}
\SIMBADobj{SDSS J081828.98+313153.0}
\SIMBADobj{SDSS J082518.86+032927.8}
\SIMBADobj{SDSS J082547.00+411900.0}
\SIMBADobj{SDSS J083203.98+142942.3}
\SIMBADobj{WD 0836+404}
\SIMBADobj{SDSS J084054.14+145709.0}
\SIMBADobj{SDSS J084220.73+370701.7}
\SIMBADobj{SDSS J084314.05+043131.6}
\SIMBADobj{SDSS J084746.82+451006.3}
\SIMBADobj{SDSS J085128.17+060551.1}
\SIMBADobj{WD 0850+002}
\SIMBADobj{SDSS J085507.29+063540.9}
\SIMBADobj{SDSS J085648.33+185804.9}
\SIMBADobj{SDSS J090041.08+190714.3}
\SIMBADobj{WD 0858+363}
\SIMBADobj{SDSS J090231.76+183554.9}
\SIMBADobj{SDSS J090624.26-002428.2}
\SIMBADobj{SDSS J091118.42+031045.1}
\SIMBADobj{SDSS J091312.74+403628.7}
\SIMBADobj{SDSS J091635.07+385546.2}
\SIMBADobj{SDSS J091731.00+092638.1}
\SIMBADobj{SDSS J092329.81+012020.0}
\SIMBADobj{WD 0921+354}
\SIMBADobj{SDSS J092511.60+050932.4}
\SIMBADobj{SDSS J093944.89+560940.2}
\SIMBADobj{SDSS J094000.27+005207.1}
\SIMBADobj{SDSS J094213.13+573342.5}
\SIMBADobj{SDSS J094917.04-000023.6}
\SIMBADobj{HS 0951+1312}
\SIMBADobj{HS 0952+1816}
\SIMBADobj{SDSS J095833.13+013049.3}
\SIMBADobj{SDSS J095936.96+023828.4}
\SIMBADobj{SDSS J100238.58+581835.9}
\SIMBADobj{SDSS J100718.26+524519.8}
\SIMBADobj{SDSS J101519.65+595430.5}
\SIMBADobj{SDSS J101548.01+030648.4}
\SIMBADobj{WD 1039+412}
\SIMBADobj{WD 1047+335}
\SIMBADobj{SDSS J105449.87+530759.1}
\SIMBADobj{SDSS J105612.32-000621.7}
\SIMBADobj{SDSS J110525.70-161328.3}
\SIMBADobj{WD 1103+015}
\SIMBADobj{WD 1116+026}
\SIMBADobj{SDSS J112221.10+035822.4}
\SIMBADobj{SDSS J112542.84+034506.3}
\SIMBADobj{WD 1126-222}
\SIMBADobj{SDSS J113604.01-013658.1}
\SIMBADobj{WD 1137+423}
\SIMBADobj{WD 1149+057}
\SIMBADobj{WD 1150-153}
\SIMBADobj{SDSS J115707.43+055303.6}
\SIMBADobj{SDSS J120054.55-025107.0}
\SIMBADobj{WD 1159+803}
\SIMBADobj{SDSS J121628.55+092246.4}
\SIMBADobj{WD 1215+009}
\SIMBADobj{SDSS J122229.57-024332.5}
\SIMBADobj{WD 1236-495}
\SIMBADobj{HS 1249+0426}
\SIMBADobj{WD 1253+024}
\SIMBADobj{SDSS J125710.50+012422.9}
\SIMBADobj{WD 1258+013}
\SIMBADobj{WD 1307+354}
\SIMBADobj{WD 1307-017}
\SIMBADobj{WD 1321+013}
\SIMBADobj{WD 1334+013}
\SIMBADobj{SDSS J133831.74-002328.0}
\SIMBADobj{WD 1342-237}
\SIMBADobj{SDSS J134550.93-005536.5}
\SIMBADobj{WD 1349+552}
\SIMBADobj{WD 1350+656}
\SIMBADobj{SDSS J135459.89+010819.3}
\SIMBADobj{SDSS J135531.03+545404.5}
\SIMBADobj{WD 1401-147}
\SIMBADobj{EQ J1408+0445}
\SIMBADobj{SDSS J141708.81+005827.2}
\SIMBADobj{WD 1422+095}
\SIMBADobj{HE 1429-0343}
\SIMBADobj{WD 1425-811}
\SIMBADobj{SDSS J144330.93+013405.8}
\SIMBADobj{SDSS J150207.02-000147.1}
\SIMBADobj{SDSS J152403.25-003022.9}
\SIMBADobj{WD 1526+558}
\SIMBADobj{HS 1531+7436}
\SIMBADobj{SDSS J153332.96-020655.7}
\SIMBADobj{WD 1541+650}
\SIMBADobj{WD 1559+369}
\SIMBADobj{WD 1607+205}
\SIMBADobj{SDSS J161218.08+083028.1}
\SIMBADobj{SDSS J161737.63+432443.8}
\SIMBADobj{WD 1616-002}
\SIMBADobj{HS 1625+1231}
\SIMBADobj{SDSS J164115.61+352140.6}
\SIMBADobj{WD 1647+591}
\SIMBADobj{SDSS J165020.53+301021.2}
\SIMBADobj{WD 1659+662}
\SIMBADobj{SDSS J170055.38+354951.1}
\SIMBADobj{SDSS J171113.01+654158.3}
\SIMBADobj{WD 1714-547}
\SIMBADobj{SDSS J172428.42+583539.0}
\SIMBADobj{SDSS J173235.19+590533.4}
\SIMBADobj{HS 1824+6000}
\SIMBADobj{WD 1855+338}
\SIMBADobj{KIC 4552982}
\SIMBADobj{SDSS J191719.16+392718.8}
\SIMBADobj{KIC 8293193}
\SIMBADobj{KIC 11911480}
\SIMBADobj{WD 1935+276}
\SIMBADobj{WD 1950+250}
\SIMBADobj{WD 1959+059}
\SIMBADobj{WD 2102+233}
\SIMBADobj{SDSS J212808.49-000750.8}
\SIMBADobj{SDSS J213530.32-074330.7}
\SIMBADobj{SDSS J214723.73-001358.4}
\SIMBADobj{WD 2148+539}
\SIMBADobj{WD 2148-291}
\SIMBADobj{WD 2151-077}
\SIMBADobj{SDSS J215628.26-004617.2}
\SIMBADobj{SDSS J215905.52+132255.7}
\SIMBADobj{SDSS J220830.02+065448.7}
\SIMBADobj{SDSS J220915.84-091942.5}
\SIMBADobj{SDSS J221458.37-002511.7}
\SIMBADobj{SDSS J223135.71+134652.8}
\SIMBADobj{SDSS J223726.86-010110.9}
\SIMBADobj{WD 2254+126}
\SIMBADobj{WD 2303+242}
\SIMBADobj{SDSS J230726.66-084700.3}
\SIMBADobj{WD 2326+049}
\SIMBADobj{SDSS J233458.71+010303.1}
\SIMBADobj{WD 2336-079}
\SIMBADobj{WD 2347+128}
\SIMBADobj{SDSS J235040.72-005430.9}
\SIMBADobj{WD 2348-244}
\SIMBADobj{SDSS J005208.42-005134.7}
\SIMBADobj{SDSS J011123.90+000935.2}
\SIMBADobj{SDSS J020351.29+004025.0}
\SIMBADobj{SDSS J082429.01+172345.4}
\SIMBADobj{SDSS J104358.59+060320.9}
\SIMBADobj{SDSS J111710.53-125540.9}
\SIMBADobj{SDSS J113655.18+040952.6}
\SIMBADobj{SDSS J111215.82+111745.0}
\SIMBADobj{SDSS J151826.68+065813.2}
\SIMBADobj{SDSS J161431.28+191219.4}
\SIMBADobj{SDSS J161831.69+385415.1}
\SIMBADobj{PSR J1738+0333}
\SIMBADobj{SDSS J184037.78+642312.3}
\SIMBADobj{SDSS J222859.93+362359.6}
\SIMBADobj{SDSS J010415.99+144857.4}
\SIMBADobj{SDSS J023520.02-093456.3}
\SIMBADobj{WD 1017-138}

\IBVSabs{We aimed at collecting all known white dwarf pulsators with hydrogen-dominated atmospheres }
\IBVSabs{and list their main photometric and atmospheric parameters together with their pulsation }
\IBVSabs{periods and amplitudes observed at different epochs. For this purpose, we explored the }
\IBVSabs{pulsating white dwarf related literature with the systematic use of the SIMBAD and the }
\IBVSabs{NASA's Astrophysics Data System (ADS) databases. We summarized our results in four tables }
\IBVSabs{listing seven ZZ~Ceti stars in detached white dwarf plus main-sequence binaries, seven } 
\IBVSabs{extremely low-mass DA pulsators, three hot DAVs and 180 ZZ~Ceti stars. }

\begintext

\section{Introduction}
The tradition of collecting and publishing the main photometric and physical
parameters of pulsating white dwarf (WD) stars, together with their observed pulsation
periods, amplitudes and their references, goes back to 1995. Bradley (1995)
collected this information on the ZZ~Ceti (or DAV), V777 Her (or DBV), interacting
binary white dwarf (IBWD) stars, and also on the pulsating PG~1159-type (or DOV) and planetary 
nebula nucleus variable (PNNV) stars known at that time.
Two updates followed this paper published in 1998 and 2000, respectively 
(Bradley 1998, 2000).
After that time, online-only updates were presented for the white dwarf community in 
2001\footnote{http://astro.if.ufrgs.br/wdtab.htm; by Kepler de Souza Oliveira Filho}, 
in 2005\footnote{http://whitedwarf.org/tables/; webpage of the White Dwarf Research Corporation} 
and in 2010\footnote{http://astro.if.ufrgs.br/zzceti.htm; by Kepler de Souza Oliveira Filho}. 
The last source lists the ZZ~Ceti stars only (155 items).

In the meantime, dividing the hot (pre-)white dwarf pulsators into DOV and PNNV classes
have become obsolete, and now the denomination GW Vir stars is used instead for all the post-AGB
stars showing nonradial \textit{g}-mode pulsations (Quirion et~al. 2007).
Nevertheless, new groups of pulsating white dwarfs have been discovered, like the
extremely low-mass DA pulsators (ELM-DAVs), the hot DAV stars, the DQV stars and
the pulsating, mixed-atmosphere, extremely low-mass white dwarf precursors 
(pre-ELM WD variables). ZZ~Ceti variables in detached white dwarf plus main-sequence (MS)
binaries have also become known. Considering the new groups, the newly discovered members 
of the `classical' ZZ~Ceti, V777 Her and GW Vir groups, 
and also the new observational results on the formerly known pulsators, the 
update of the white dwarf data tables is now appropriate.

In this paper, taking account of the relatively large number of white dwarf variables
and the considerable observational information on them, we focus on their most populated
subgroup, that is, on the variables with hydrogen-dominated atmospheres only. Thus, we 
collected the main stellar parameters and pulsational properties
with references of the `classical' ZZ~Ceti stars, the ZZ~Ceti variables in detached
WD plus MS binaries, the ELM-DAV and hot DAV stars.

\section{Data collection and structure of the data tables}
\label{sect:datacoll}

Table~\ref{dams} lists the ZZ~Ceti variables in detached white dwarf plus main-sequence
binaries. This list of seven variables is based on the paper of Pyrzas et~al. (2015).

Table~\ref{elmda} summarizes the observational results on the seven extremely low-mass DA
pulsators presented in the papers of Hermes et~al. (2012, 2013b,d); Bell et~al. (2015b) 
and Kilic et~al. (2015).

The hot DAV stars are listed in Table~\ref{hotdav}, a new group consisting of three members
discovered by the work of Kurtz et~al. (2008, 2013).

Finally, we list the members of the most populated subgroup of white dwarf pulsators
with hydrogen atmospheres, the `classical' ZZ~Ceti stars, in Table~\ref{zzceti}.
These objects can be treated as products of single-star evolution, in contrast to
the ELM-DAV or ZZ~Ceti stars in WD+MS binaries. We started the data collection
with the 136 DAVs listed and refereed in the review paper of Fontaine \& Brassard (2008).
We then complemented this list with the ZZ~Ceti stars reported by
Stobie et~al. (1997); Castanheira et~al. (2010); Hermes et~al. (2011); Sayres et~al. (2012);
Kepler et~al. (2012); Castanheira et~al. (2013); Hermes et~al. (2013a); Greiss et~al. (2014); 
Green et~al. (2015); Gentile Fusillo et~al. (2016); Greiss et~al. (2016) and Bell et~al. (2016).
We closed the data collection with this last paper published on arxiv.org on 7th July, 2016.
Altogether, 180 ZZ~Ceti stars are listed in Table~\ref{zzceti}.

All tables follow the same structure specified below.

\begin{itemize}
 \item Identifiers: the first column shows the star's identifier in WD~(J)HHMM$\pm$DDM(M) format,
 while the second column shows another identifier used in the literature or can be used to
 identify the object in the SIMBAD database (Wenger et~al. 2000). The identifiers
 in parentheses are not recognized by SIMBAD, but used in the literature.
 \item Third and fourth columns: right ascension (RA) and declination (DEC) in the equatorial 
 coordinate system (epoch J2000.0) from the SIMBAD database. The objects are arranged 
 according to increasing right ascensions.
 \item Fifth and sixth columns: effective temperature ($T_{\mathrm{eff}}$) and surface gravity 
 ($\mathrm{log}\,g$) values. In the case of ZZ~Ceti stars in Table~\ref{zzceti}, most of the 
 objects can be found either in the database of Gianninas et~al. (2011) or Tremblay et~al. (2011) 
 with their spectroscopic atmospheric parameters determined by one-dimensional model atmospheres
 and by the use of the ML2/$\alpha$=0.8 version of the mixing-length theory. 
 We corrected these $T_{\mathrm{eff}}$ and $\mathrm{log}\,g$ values according to the findings of 
 Tremblay et~al. (2013) based on radiation-hydrodynamics three-dimensional 
 simulations of convective DA stellar atmospheres. We denoted the resulting values with
 \textit{G} or \textit{T} in superscript at the corresponding effective temperatures referring to the 
 source of the original atmospheric parameters. In all the other cases, the source of the $T_{\mathrm{eff}}$ 
 and $\mathrm{log}\,g$ values is the paper referred to in the last column, which is practically the 
 paper reporting the discovery of the given pulsator.
 \item Seventh column: the \textit{V} magnitude of the star in the SIMBAD database. If there is no
 such data in SIMBAD, we list its \textit{g} magnitude, or, in absence of both \textit{V} and \textit{g}
 magnitudes, its brightness in \textit{B} band.
 \item Eight and ninth (last) columns: pulsation period (in second) with Fourier amplitude (in milli-modulation amplitude, mma) 
 values arranged according to increasing periods, and the corresponding references. We applied the 
 \begin{equation*}
 1\,\mathrm{mma}=1/1.086\,\mathrm{mmag}=0.1\%=1\,\mathrm{ppt} 
 \end{equation*}
 conversion to convert the amplitudes published if necessary.
 
 Our data collection strategy was to use the NASA's Astrophysics Data System (ADS) to search for 
 publications referring to a given object and looked for papers publishing pulsation periods and their Fourier
 amplitudes. In some instances, only period values were presented. In such cases, we abstained from making 
 estimates on the amplitudes by the Fourier spectra presented (if any), thus we listed the periods only. 
 
 We aimed at collecting linearly
 independent pulsation frequencies alone, thus, if a frequency was denoted as a combination term in the 
 literature, we did not add it to the period list. Such cases are indicated in parentheses by the following remarks after 
 the periods and amplitudes: `+C' --  there are 
 additional linear combinations (including harmonics) reported; `+SH' -- additional subharmonics ($\sim n/2f_i$) present in the data.
 `+R' denotes that frequency components, which may be results of rotational splitting are also detected.
 In addition, the `iR' remark means that the period list contains rotationally split frequencies, 
 while `iC?' denotes that our list may contains combination terms.
\end{itemize}

\section{What is new?}

One of the most conspicuous improvements we made comparing to the previous versions of white
dwarf data tables is the completion of the object list with newly discovered pulsators both
in the formerly known group of classical ZZ~Ceti stars and in the newly established groups
of hydrogen-atmosphere white dwarf pulsators.

Another relevant improvement concerns the periods listed. The authors restricted to the presentation
of one set of periods, or just a representative period range, per object in the previously published white dwarf data tables. 
In contrast, we attempted to collect all different
period lists existing in the literature for an object, that is, observational results from different epochs.
We also emphasize this choice with the title selection of our catalogue `White Dwarf Period Tables' instead of the 
`White Dwarf Data Tables' used previously. This also implies that we present the almost complete bibliography
of the observations of hydrogen-atmosphere pulsating white dwarf stars starting from 1968.

The different periods observed at different epochs can be the result e.g. of the different
lengths and qualities of the data sets analysed, however, especially in the case of the ZZ~Ceti stars being 
close to the red edge of the instability domain, short time-scale variations in the amplitudes of the excited modes are common.
That is, pulsation modes never seen can be excited to an observable level, while others can vanish from one observational run 
to another. Eventually, comparing the different sets of periods from different runs, it can result in a more complete set of 
pulsation periods, which is essential for detailed asteroseismic investigations.

Smaller, but also significant improvements that we checked and corrected
the star identifiers if it was necessary, in order to publish at least one identifier per star which can be found in the 
SIMBAD database. In essence, this affected the Sloan Digital Sky Survey (SDSS) identifiers, where the use of the correct 
format, including all the necessary decimals is crucial.
We also note that in many cases the WD~(J)HHMM$\pm$DDM(M) identifiers used in the literature are not found in SIMBAD,
thus, especially in the publication of a new discovery, we recommend the indication of another identifier existing
in the database (if any), or at least the equatorial coordinates of the object to make the identification of the new 
pulsator clear and easy.
At last, as we mentioned in Sect.~\ref{sect:datacoll}, we revised and updated the effective temperature and surface gravity values
of the classical ZZ~Ceti stars.


\begin{landscape}
\vskip 5mm
\begin{table}
\caption{ZZ~Ceti stars in detached white dwarf plus main-sequence binaries.}
\vskip 5mm
\begin{center}
\footnotesize
\begin{tabular}{rrrrlp{6mm}lrr}
\hline
\multicolumn{2}{c}{Identifiers} & RA & DEC & \multicolumn{1}{c}{$T_{\mathrm{eff}}$} & $\mathrm{log}\,g$ & \multicolumn{1}{c}{\textit{V}} & Period/Amplitude & References\\
WD (SDSS) & Other & (h m s) & (\deg\,\arcm\,\arcs) & \multicolumn{1}{c}{(K)} & (dex) & \multicolumn{1}{c}{(mag)} & (s/mma) & \\
\hline
0049-011 & SDSS J005208.42-005134.7 & 00 52 08 & $-$00 51 35 & 12\,300 & 8.46 & 18.3 (g) & 1077.2/4.0; 1093.8/4.0 & Pyrzas et~al. (2015)\\ 
\\
(J0111+000) & SDSS J011123.90+000935.2 & 01 11 24 & +00 09 35 & 12\,320 & 7.50 & 18.4 (g) & 366.5/9.1; 510.2/18.9; 583.2/16.3; & Pyrzas et~al. (2015)\\
 & & & & & & & 631.6/28.0; 883.6/15.9 & \\
\\
0201+004 & SDSS J020351.29+004025.0 & 02 03 51 & +00 40 25 & 10\,790 & 8.17 & 19.4 (g) & 398.8/11.9; 683.9/15.2; 957.0/38.5 & Pyrzas et~al. (2015)\\
\\
(J0824+1723) & SDSS J082429.01+172345.4 & 08 24 29 & +17 23 45 & 11\,430 & 8.21 & 18.3 (g) & 513.9/9.2; 623.7/20.4; 807.2/13.9; & Pyrzas et~al. (2015)\\
 & & & & & & & 987.9/12.0 & \\
\\
(1043+0603) & SDSS J104358.59+060320.9 & 10 43 59 & +06 03 21 & 11\,170 & 8.19 & 18.7 (g) & 164.9/6.6; 292.5/5.3; 324.7/28.8; & Pyrzas et~al. (2015)\\
 & & & & & & & 634.4/8.5 & \\
\\
(J1117-1255) & SDSS J111710.53-125540.9 & 11 17 11 & $-$12 55 41 & 11\,300 & 8.29 & 19.7 (g) & 835.8/21.9 & Pyrzas et~al. (2015)\\
\\
(J1136+0409) & SDSS J113655.18+040952.6 & 11 36 55 & +04 09 53 & 11\,700 & 7.99 & 17.1 & 182.2/4.4; 276.5/8.3 & Pyrzas et~al. (2015)\\
\\
 & & & & & & & 160.8/0.4; 162.2/1.2; 163.7/0.4; & Hermes et~al. (2015)\\
 & & & & & & & 181.3/1.8; 201.8/0.5; 274.9/3.5; & \\
 & & & & & & & 279.4/2.3; 284.2/0.2; 337.7/0.8; & \\
 & & & & & & & 351.1/0.7; 395.9/0.4; 474.5/0.4 (iR) & \\
\hline
\label{dams}
\end{tabular}
\end{center}
\end{table}
\end{landscape}
\vskip 5mm


\begin{landscape}
\vskip 5mm
\begin{table}
\caption{Extremely low-mass DA pulsators.}
\vskip 5mm
\begin{center}
\footnotesize
\begin{tabular}{rrrrlp{6mm}lrr} 
\hline
\multicolumn{2}{c}{Identifiers} & RA & DEC & \multicolumn{1}{c}{$T_{\mathrm{eff}}$} & $\mathrm{log}\,g$ & \multicolumn{1}{c}{\textit{V}} & Period/Amplitude & References\\
WD (SDSS) & Other & (h m s) & (\deg\,\arcm\,\arcs) & \multicolumn{1}{c}{(K)} & (dex) & \multicolumn{1}{c}{(mag)} & (s/mma) & \\
\hline
(J1112+1117) & SDSS J111215.82+111745.0 & 11 12 16 & +11 17 45 & 9\,590 & 6.36 & 16.3 & 107.6/0.4; 134.3/0.4; 1792.9/3.3; & Hermes et~al. (2013d)\\
 & & & & & & & 1884.6/4.7; 2258.5/7.5; & \\
 & & & & & & & 2539.7/6.8; 2855.7/3.6 & \\
\\
(J1518+0658) & SDSS J151826.68+065813.2 & 15 18 27 & +06 58 13 & 9\,900 & 6.80 & 17.6 (g) & 1335.3/13.6; 1956.4/18.1; & Hermes et~al. (2013d)\\
 & & & & & & & 2134.0/14.2; 2268.2/21.6; & \\
 & & & & & & & 2714.3/21.6; 2799.1/35.4; & \\
 & & & & & & & 3848.2/15.7 & \\
\\
 & & & & & & & 1318.8/12.9; 1956.3/18.1; & Hermes et~al. (2013d)\\
 & & & & & & & 2210.0/19.9; 2413.1/15.6; & \\
 & & & & & & & 2796.0/41.1; 2802.8/26.4; & \\
 & & & & & & & 3683.7/17.7 & \\
\\
(J1614+1912) & SDSS J161431.28+191219.4 & 16 14 31 & +19 12 19 & 8\,880 & 6.66 & 16.4 (g) & 1184.1/3.2; 1262.7/5.9 & Hermes et~al. (2013b)\\
\\
(J1618+3854) & SDSS J161831.69+385415.1 & 16 18 32 & +38 54 15 & 9\,140 & 6.83 & 19.7 (g) & 2543.0/16.0; 4935.2/56.3; & Bell et~al. (2015b)\\
 & & & & & & &  6125.9/25.5 & \\
\\
-- & PSR J1738+0333 & 17 38 54 & +03 33 11 & 9\,130 & 6.55 &  21.3 & 1788.0/12.7; 2656.0/11.5; & Kilic et~al. (2015)\\
 & & & & & & & 3057.0/12.2 & \\
\\
(J1840+6423) & SDSS J184037.78+642312.3 & 18 40 38 & +64 23 12 & 9\,100 & 6.22 & 18.9 (g) & 1578.6/29.5; 2376.0/48.8; & Hermes et~al. (2012)\\
 & & & & & & & 4445.9/65.9 & \\
\\
(J2228+3623) & SDSS J222859.93+362359.6 & 22 28 60 & +36 24 00 & 7\,870 & 6.03 & 16.9 (g) & 3254.5/2.3; 4178.3/6.3; 6234.9/1.9 & Hermes et~al. (2013b)\\
\hline
\label{elmda}
\end{tabular}
\end{center}
\end{table}
\end{landscape}
\vskip 5mm


\begin{landscape}
\vskip 5mm
\begin{table}
\caption{Hot DAV stars.}
\vskip 5mm
\begin{center}
\footnotesize
\begin{tabular}{rrrrlp{6mm}lrr}
\hline
\multicolumn{2}{c}{Identifiers} & RA & DEC & \multicolumn{1}{c}{$T_{\mathrm{eff}}$} & $\mathrm{log}\,g$ & \multicolumn{1}{c}{\textit{V}} & Period/Amplitude & References\\
WD & Other & (h m s) & (\deg\,\arcm\,\arcs) & \multicolumn{1}{c}{(K)} & (dex) & \multicolumn{1}{c}{(mag)} & (s/mma) & \\
\hline
0101+145 & SDSS J010415.99+144857.4 & 01 04 16 & +14 48 57 & 29\,980 & 7.38 & 18.8 (g) & 159.0/4.0 & Kurtz et~al. (2008)\\ 
\\
0232-097A & SDSS J023520.02-093456.3 & 02 35 20 & $-$09 34 56 & 30\,110 & 7.30 & 17.8 (g) & 705.0/15.6 & Kurtz et~al. (2008)\\
\\
1017-138 & -- & 10 19 52 & $-$14 07 34 & 32\,600 & 7.80 & 14.6 & 624.0/1.0 & Kurtz et~al. (2013)\\
\hline
\label{hotdav}
\end{tabular}
\end{center}
\end{table}
\end{landscape}
\vskip 5mm


\begin{landscape}
\footnotesize
\begin{longtable}{rrrrp{8mm}p{4mm}p{13mm}rr}
\caption{List of ZZ~Ceti stars. \label{zzceti}}\\
\hline
\multicolumn{2}{c}{Identifiers} & RA & DEC & \multicolumn{1}{c}{$T_{\mathrm{eff}}$} & $\mathrm{log}\,g$ & \multicolumn{1}{c}{\textit{V}} & Period/Amplitude & References\\
WD & Other & (h m s) & (\deg\,\arcm\,\arcs) & \multicolumn{1}{c}{(K)} & (dex) & \multicolumn{1}{c}{(mag)} & (s/mma) & \\
\hline
\endfirsthead
\caption{continued.}\\
\hline
\multicolumn{2}{c}{Identifiers} & RA & DEC & \multicolumn{1}{c}{$T_{\mathrm{eff}}$} & $\mathrm{log}\,g$ & \multicolumn{1}{c}{\textit{V}} & Period/Amplitude & References\\
WD & Other & (h m s) & (\deg\,\arcm\,\arcs) & \multicolumn{1}{c}{(K)} & (dex) & \multicolumn{1}{c}{(mag)} & (s/mma) & \\
\hline
\endhead
\hline
\endfoot
\hline
\endlastfoot
(J0000-0046) & SDSS J000006.75-004654.0 & 00 00 07 & $-$00 46 54 & 10\,620$^{T}$ & 8.18 & 18.8 (g) & 584.8/15.9; 601.4/9.0; 611.4/23.0 & Castanheira et~al. (2007)\\
\\
(J0018+0031) & SDSS J001836.11+003151.1 & 00 18 36 & 00 31 53 & 11\,530$^{T}$ & 8.04 & 17.4 (g) & 257.9/5.8 & Mullally et~al. (2005)\\
\\
 & & & & & & & 149.9/3.7; 257.3/6.0 & Castanheira \& Kepler (2009)\\
\\
0016-258 & MCT 0016-2553 & 00 18 45 & $-$25 36 42 & 11\,060$^{G}$ & 8.06 & 16.1 & 1152.4/8.1 & Gianninas et~al. (2006)\\
\\
\end{longtable}
\blfootnote{{\bf Notes.} $^{(G)}$ The effective temperature ($T_{\mathrm{eff}}$) and surface gravity ($\mathrm{log}\,g$) values are provided by 
  Gianninas et~al. (2011), and then corrected according to the results of Tremblay et~al. (2013).
  $^{(T)}$ The effective temperature ($T_{\mathrm{eff}}$) and surface gravity ($\mathrm{log}\,g$) values are provided by 
  Tremblay et~al. (2011), and then corrected according to the results of Tremblay et~al. (2013).}
\end{landscape}
\vskip 5mm

\vskip 1.5 mm

\emph{Acknowledgements:} The financial support of the Hungarian National Research, Development and Innovation Office 
(NKFIH) grant K-115709 is acknowledged.
\'A.S. was supported by the J\'anos Bolyai Research Scholarship of the Hungarian Academy of Sciences. 
This research has made use of the SIMBAD database, operated at CDS, Strasbourg, France, and 
NASA's Astrophysics Data System Bibliographic Services.

\vskip -2 mm

\references

Bell, K.~J., Hermes, J.~J., Bischoff-Kim, A., et~al., 2015a, {\it \apj}, {\bf 809}, 14 \BIBCODE{2015ApJ...809...14B}

Bell, K.~J., Hermes, J.~J., Montgomery, M.~H., et~al., 2016, {\it ArXiv e-prints}, [arXiv]1607.01392 \BIBCODE{2016arXiv160701392B}

Bell, K.~J., Kepler, S.~O., Montgomery, M.~H., et~al., 2015b, {\it Astronomical Society of the Pacific Conference Series}, 
{\bf 493}, 19th European Workshop on White Dwarfs, ed. P.~Dufour, P.~Bergeron, \& G.~Fontaine, p. 217 \BIBCODE{2015ASPC..493..217B}

Bergeron, P., Fontaine, G., Bill\`eres, M., Boudreault, S., \& Green, E.~M., 2004, {\it \apj}, {\bf 600}, 404 \BIBCODE{2004ApJ...600..404B}

Bergeron, P., Fontaine, G., Brassard, P., et~al., 1993, {\it \aj}, {\bf 106}, 1987 \BIBCODE{1993AJ....106.1987B}

Bergeron, P. \& McGraw, J.~T., 1990, {\it \apjl}, {\bf 352}, L45 \BIBCODE{1990ApJ...352L..45B}

Bogn\'ar, Zs., Papar\'o, M., Bradley, P.~A., \& Bischoff-Kim, A., 2009, {\it \mnras}, {\bf 399}, 1954 \BIBCODE{2009MNRAS.399.1954B}

Bogn\'ar, Zs., Papar\'o, M., Moln\'ar, L., et~al., 2016, {\it \mnras}, {\bf 461}, 4059 \BIBCODE{2016MNRAS.461.4059B}

Bogn\'ar, Zs., Papar\'o, M., Steininger, B., \& Vir\'aghalmy, G., 2007, {\it Communications in Asteroseismology}, {\bf 150}, 251 \BIBCODE{2007CoAst.150..251B}

Bradley, P.~A., 1995, {\it Baltic Astronomy}, {\bf 4}, 536 \BIBCODE{1995BaltA...4..536B}

Bradley, P.~A., 1998, {\it Baltic Astronomy}, {\bf 7}, 355 \BIBCODE{1998BaltA...7..355B}

Bradley, P.~A., 2000, {\it Baltic Astronomy}, {\bf 9}, 485 \BIBCODE{2000BaltA...9..485B}

Castanheira, B.~G. \& Kepler, S.~O., 2008, {\it \mnras}, {\bf 385}, 430 \BIBCODE{2008MNRAS.385..430C}

Castanheira, B.~G. \& Kepler, S.~O., 2009, {\it \mnras}, {\bf 396}, 1709 \BIBCODE{2009MNRAS.396.1709C}

Castanheira, B.~G., Kepler, S.~O., Costa, A.~F.~M., et~al., 2007, {\it \aap}, {\bf 462}, 989 \BIBCODE{2007A&A...462..989C}

Castanheira, B.~G., Kepler, S.~O., Kleinman, S.~J., Nitta, A., \& Fraga, L., 2010, {\it \mnras}, {\bf 405}, 2561 \BIBCODE{2010MNRAS.405.2561C}

Castanheira, B.~G., Kepler, S.~O., Kleinman, S.~J., Nitta, A., \& Fraga, L., 2013, {\it \mnras}, {\bf 430}, 50 \BIBCODE{2013MNRAS.430...50C}

Castanheira, B.~G., Kepler, S.~O., Moskalik, P., et~al., 2004, {\it \aap}, {\bf 413}, 623 \BIBCODE{2004A&A...413..623C}

Castanheira, B.~G., Kepler, S.~O., Mullally, F., et~al., 2006, {\it \aap}, {\bf 450}, 227 \BIBCODE{2006A&A...450..227C}

Chen, Y.~H. \& Li, Y., 2014, {\it \mnras}, {\bf 443}, 3477 \BIBCODE{2014MNRAS.443.3477C}

Dolez, N., 1998, {\it Baltic Astronomy}, {\bf 7}, 153 \BIBCODE{1998BaltA...7..153D}

Dolez, N., Vauclair, G., \& Chevreton, M., 1983, {\it \aap}, {\bf 121}, L23 \BIBCODE{1983A&A...121L..23D}

Dolez, N., Vauclair, G., Kleinman, S.~J., et~al., 2006, {\it \aap}, {\bf 446}, 237 \BIBCODE{2006A&A...446..237D}

Dolez, N., Vauclair, G., Xiao-Bin, Z., Chevreton, M., \& Handler, G., 1999, {\it Astronomical Society 
of the Pacific Conference Series}, {\bf 169}, 11th European Workshop on White Dwarfs, ed. S.-E. Solheim \& E.~G.
Meistas, p. 129 \BIBCODE{1999ASPC..169..129D}

Fitch, W.~S., 1973, {\it \apjl}, {\bf 181}, L95 \BIBCODE{1973ApJ...181L..95F}

Fontaine, G., Bergeron, P., Bill\`eres, M., \& Charpinet, S., 2003, {\it \apj}, {\bf 591}, 1184 \BIBCODE{2003ApJ...591.1184F}

Fontaine, G., Bergeron, P., Brassard, P., Bill\`eres, M., \& Charpinet, S., 2001, {\it \apj}, {\bf 557}, 792 \BIBCODE{2001ApJ...557..792F}

Fontaine, G. \& Brassard, P., 2008, {\it \pasp}, {\bf 120}, 1043 \BIBCODE{2008PASP..120.1043F}

Fontaine, G., Lacombe, P., McGraw, J.~T., et~al., 1980, {\it \apj}, {\bf 239}, 898 \BIBCODE{1980ApJ...239..898F}

Fontaine, G. \& Wesemael, F., 1984, {\it \aj}, {\bf 89}, 1728 \BIBCODE{1984AJ.....89.1728F}

Fontaine, G., Wesemael, F., Bergeron, P., et~al., 1985, {\it \apj}, {\bf 294}, 339 \BIBCODE{1985ApJ...294..339F}

Fu, J.-N., Dolez, N., Vauclair, G., et~al., 2013, {\it \mnras}, {\bf 429}, 1585 \BIBCODE{2013MNRAS.429.1585F}

Fu, J.-N., Vauclair, G., Dolez, N., Jiang, S.-Y., \& Chevreton, M., 2002, {\it Astronomical Society of 
the Pacific Conference Series}, {\bf 259}, IAU Colloq. 185: Radial and Nonradial Pulsations as Probes of Stellar Physics, 
ed. C.~Aerts, T.~R. Bedding, \& J.~Christensen-Dalsgaard, p. 378 \BIBCODE{2002ASPC..259..378F}

Gentile Fusillo, N.~P., Hermes, J.~J., \& G\"ansicke, B.~T., 2016, {\it \mnras}, {\bf 455}, 2295 \BIBCODE{2016MNRAS.455.2295G}

Giammichele, N., Fontaine, G., Bergeron, P., et~al., 2015, {\it \apj}, {\bf 815}, 56 \BIBCODE{2015ApJ...815...56G}

Gianninas, A., Bergeron, P., \& Fontaine, G., 2005, {\it \apj}, {\bf 631}, 1100 \BIBCODE{2005ApJ...631.1100G}

Gianninas, A., Bergeron, P., \& Fontaine, G., 2006, {\it \aj}, {\bf 132}, 831 \BIBCODE{2006AJ....132..831G}

Gianninas, A., Bergeron, P., \& Ruiz, M.~T., 2011, {\it \apj}, {\bf 743}, 138 \BIBCODE{2011ApJ...743..138G}

Giovannini, O., Kepler, S.~O., Kanaan, A., Costa, A.~F.~M., \& Koester, D., 1998, {\it \aap}, {\bf 329}, L13 \BIBCODE{1998A&A...329L..13G}

Green, E.~M., Limoges, M.-M., Gianninas, A., et~al., 2015, {\it Astronomical Society of the Pacific 
Conference Series}, {\bf 493}, 19th European Workshop on White Dwarfs, ed. P.~Dufour, P.~Bergeron, \& G.~Fontaine, p. 237 \BIBCODE{2015ASPC..493..237G}

Greiss, S., G\"ansicke, B.~T., Hermes, J.~J., et~al., 2014, {\it \mnras}, {\bf 438}, 3086 \BIBCODE{2014MNRAS.438.3086G}

Greiss, S., Hermes, J.~J., G\"ansicke, B.~T., et~al., 2016, {\it \mnras}, {\bf 457}, 2855 \BIBCODE{2016MNRAS.457.2855G}

Handler, G., Provencal, J.~L., Lendl, M., Montgomery, M.~H., \& Beck, P.~G., 2008a, {\it Communications in Asteroseismology}, {\bf 156}, 18 \BIBCODE{2008CoAst.156...18H}

Handler, G. \& Romero-Colmenero, E., 2001, {\it Astronomical Society of the Pacific Conference Series}, {\bf 226}, 
12th European Workshop on White Dwarfs, ed. J.~L. Provencal, H.~L. Shipman, J.~MacDonald, \& S.~Goodchild, p. 313 \BIBCODE{2001ASPC..226..313H}

Handler, G., Romero-Colmenero, E., \& Montgomery, M.~H., 2002, {\it \mnras}, {\bf 335}, 399 \BIBCODE{2002MNRAS.335..399H}

Handler, G., Romero-Colmenero, E., Provencal, J.~L., et~al., 2008b, {\it \mnras}, {\bf 388}, 1444 \BIBCODE{2008MNRAS.388.1444H}

Hermes, J.~J., Charpinet, S., Barclay, T., et~al., 2014, {\it \apj}, {\bf 789}, 85 \BIBCODE{2014ApJ...789...85H}

Hermes, J.~J., G\"ansicke, B.~T., Bischoff-Kim, A., et~al., 2015, {\it \mnras}, {\bf 451}, 1701 \BIBCODE{2015MNRAS.451.1701H}

Hermes, J.~J., Kepler, S.~O., Castanheira, B.~G., et~al., 2013a, {\it \apjl}, {\bf 771}, L2 \BIBCODE{2013ApJ...771L...2H}

Hermes, J.~J., Montgomery, M.~H., Gianninas, A., et~al., 2013b, {\it \mnras}, {\bf 436}, 3573 \BIBCODE{2013MNRAS.436.3573H}

Hermes, J.~J., Montgomery, M.~H., Mullally, F., Winget, D.~E., \& Bischoff-Kim, A., 2013c, {\it \apj}, {\bf 766}, 42 \BIBCODE{2013ApJ...766...42H}

Hermes, J.~J., Montgomery, M.~H., Winget, D.~E., et~al., 2013d, {\it \apj}, {\bf 765}, 102 \BIBCODE{2013ApJ...765..102H}

Hermes, J.~J., Montgomery, M.~H., Winget, D.~E., et~al., 2012, {\it \apjl}, {\bf 750}, L28 \BIBCODE{2012ApJ...750L..28H}

Hermes, J.~J., Mullally, F., Ostensen, R.~H., et~al., 2011, {\it \apjl}, {\bf 741}, L16 \BIBCODE{2011ApJ...741L..16H}

Hesser, J.~E., Lasker, B.~M., \& Neupert, H.~E., 1976, {\it \apj}, {\bf 209}, 853 \BIBCODE{1976ApJ...209..853H}

H\"urkal, D.~\"O., Handler, G., Steininger, B.~A., \& Reed, M.~D., 2005, {\it Astronomical Society of the 
Pacific Conference Series}, {\bf 334}, 14th European Workshop on White Dwarfs, ed. D.~Koester \& S.~Moehler, p. 577 \BIBCODE{2005ASPC..334..577H}

Jordan, S., Koester, D., Vauclair, G., et~al., 1998, {\it \aap}, {\bf 330}, 277 \BIBCODE{1998A&A...330..277J}

Kanaan, A., Kepler, S.~O., Giovannini, O., \& Diaz, M., 1992, {\it \apjl}, {\bf 390}, L89 \BIBCODE{1992ApJ...390L..89K}

Kanaan, A., Kepler, S.~O., Giovannini, O., et~al., 1998, {\it Baltic Astronomy}, {\bf 7}, 183 \BIBCODE{1998BaltA...7..183K}

Kanaan, A., Nitta, A., Winget, D.~E., et~al., 2005, {\it \aap}, {\bf 432}, 219 \BIBCODE{2005A&A...432..219K}

Kanaan, A., Nitta-Kleinman, A., Winget, D.~E., et~al., 2000, {\it Baltic Astronomy}, {\bf 9}, 87 \BIBCODE{2000BaltA...9...87K}

Kepler, S.~O., 1984, {\it \apj}, {\bf 278}, 754 \BIBCODE{1984ApJ...278..754K}

Kepler, S.~O., Castanheira, B.~G., Saraiva, M.~F.~O., et~al., 2005, {\it \aap}, {\bf 442}, 629 \BIBCODE{2005A&A...442..629K}

Kepler, S.~O., Giovannini, O., Costa, A.~F.~M., et~al., 1995a, {\it Baltic Astronomy}, {\bf 4}, 238 \BIBCODE{1995BaltA...4..238K}

Kepler, S.~O., Giovannini, O., Wood, M.~A., et~al., 1995b, {\it \apj}, {\bf 447}, 874 \BIBCODE{1995ApJ...447..874K}

Kepler, S.~O., Nather, R.~E., McGraw, J.~T., \& Robinson, E.~L., 1982, {\it \apj}, {\bf 254}, 676 \BIBCODE{1982ApJ...254..676K}

Kepler, S.~O., Pelisoli, I., Pe\c canha, V., et~al., 2012, {\it \apj}, {\bf 757}, 177 \BIBCODE{2012ApJ...757..177K}

Kepler, S.~O., Robinson, E.~L., Koester, D., et~al., 2000, {\it Baltic Astronomy}, {\bf 9}, 59 \BIBCODE{2000BaltA...9...59K}

Kepler, S.~O., Robinson, E.~L., \& Nather, R.~E., 1983, {\it \apj}, {\bf 271}, 744 \BIBCODE{1983ApJ...271..744K}

Kepler, S.~O., Winget, D.~E., Nather, R.~E., et~al., 1995c, {\it Baltic Astronomy}, {\bf 4}, 221 \BIBCODE{1995BaltA...4..221K}

Kilic, M., Hermes, J.~J., Gianninas, A., \& Brown, W.~R., 2015, {\it \mnras}, {\bf 446}, L26 \BIBCODE{2015MNRAS.446L..26K}

Kleinman, S.~J., 1995, {\it Baltic Astronomy}, {\bf 4}, 270 \BIBCODE{1995BaltA...4..270K}

Kleinman, S.~J., Nather, R.~E., Winget, D.~E., et~al., 1998, {\it \apj}, {\bf 495}, 424 \BIBCODE{1998ApJ...495..424K}

Kotak, R., van Kerkwijk, M.~H., \& Clemens, J.~C., 2002, {\it \aap}, {\bf 388}, 219 \BIBCODE{2002A&A...388..219K}

Kurtz, D.~W., Shibahashi, H., Dhillon, V.~S., Marsh, T.~R., \& Littlefair, S.~P., 2008, {\it \mnras}, {\bf 389}, 1771 \BIBCODE{2008MNRAS.389.1771K}

Kurtz, D.~W., Shibahashi, H., Dhillon, V.~S., et~al., 2013, {\it \mnras}, {\bf 432}, 1632 \BIBCODE{2013MNRAS.432.1632K}

Landolt, A.~U., 1968, {\it \apj}, {\bf 153}, 151 \BIBCODE{1968ApJ...153..151L}

Lasker, B.~M. \& Hesser, J.~E., 1971, {\it \apjl}, {\bf 163}, L89 \BIBCODE{1971ApJ...163L..89L}

Li, C., Fu, J.-N., Vauclair, G., et~al., 2015, {\it \mnras}, {\bf 449}, 3360 \BIBCODE{2015MNRAS.449.3360L}

McGraw, J.~T., 1976, {\it \apjl}, {\bf 210}, L35 \BIBCODE{1976ApJ...210L..35M}

McGraw, J.~T., 1977, {\it \apjl}, {\bf 214}, L123 \BIBCODE{1977ApJ...214L.123M}

McGraw, J.~T., Fontaine, G., Lacombe, P., et~al., 1981, {\it \apj}, {\bf 250}, 349 \BIBCODE{1981ApJ...250..349M}

McGraw, J.~T. \& Robinson, E.~L., 1975, {\it \apjl}, {\bf 200}, L89 \BIBCODE{1975ApJ...200L..89M}

McGraw, J.~T. \& Robinson, E.~L., 1976, {\it \apjl}, {\bf 205}, L155 \BIBCODE{1976ApJ...205L.155M}

Metcalfe, T.~S., Montgomery, M.~H., \& Kanaan, A., 2004, {\it \apjl}, {\bf 605}, L133 \BIBCODE{2004ApJ...605L.133M}

Mukadam, A.~S., Bischoff-Kim, A., Fraser, O., et~al., 2013, {\it \apj}, {\bf 771}, 17 \BIBCODE{2013ApJ...771...17M}

Mukadam, A.~S., Kepler, S.~O., Winget, D.~E., \& Bergeron, P., 2002, {\it \apj}, {\bf 580}, 429 \BIBCODE{2002ApJ...580..429M}

Mukadam, A.~S., Kepler, S.~O., Winget, D.~E., et~al., 2003, {\it Baltic Astronomy}, {\bf 12}, 71 \BIBCODE{2003BaltA..12...71M}

Mukadam, A.~S., Montgomery, M.~H., Winget, D.~E., Kepler, S.~O., \& Clemens, J.~C., 2006, {\it \apj}, {\bf 640}, 956 \BIBCODE{2006ApJ...640..956M}

Mukadam, A.~S., Mullally, F., Nather, R.~E., et~al., 2004, {\it \apj}, {\bf 607}, 982 \BIBCODE{2004ApJ...607..982M}

Mullally, F., Thompson, S.~E., Castanheira, B.~G., et~al., 2005, {\it \apj}, {\bf 625}, 966 \BIBCODE{2005ApJ...625..966M}

Mullally, F., Winget, D.~E., Degennaro, S., et~al., 2008, {\it \apj}, {\bf 676}, 573 \BIBCODE{2008ApJ...676..573M}

Odonoghue, D., 1986, {\it NATO Advanced Science Institutes (ASI) Series C}, {\bf 169}, ed. D.~O. Gough, p. 467 \BIBCODE{1986ASIC..169..467O}

O'Donoghue, D., Warner, B., \& Cropper, M., 1992, {\it \mnras}, {\bf 258}, 415 \BIBCODE{1992MNRAS.258..415O}

Odonoghue, D.~E. \& Warner, B., 1982, {\it \mnras}, {\bf 200}, 563 \BIBCODE{1982MNRAS.200..563O}

Page, C.~G., 1972, {\it \mnras}, {\bf 159}, 25 \BIBCODE{1972MNRAS.159P..25P}

Pak\v stien\.e, E., Solheim, J.-E., Handler, G., et~al., 2011, {\it \mnras}, {\bf 415}, 1322 \BIBCODE{2011MNRAS.415.1322P}

Papar\'o, M., Bogn\'ar, Zs., Plachy, E., Moln\'ar, L., \& Bradley, P.~A., 2013, {\it \mnras}, {\bf 432}, 598 \BIBCODE{2013MNRAS.432..598P}

Patterson, J., Zuckerman, B., Becklin, E.~E., Tholen, D.~J., \& Hawarden, T., 1991, {\it \apj}, {\bf 374}, 330 \BIBCODE{1991ApJ...374..330P}

Pfeiffer, B., Vauclair, G., Dolez, N., et~al., 1996, {\it \aap}, {\bf 314}, 182 \BIBCODE{1996A&A...314..182P}

Pfeiffer, B., Vauclair, G., Dolez, N., et~al., 1995, {\it Baltic Astronomy}, {\bf 4}, 245 \BIBCODE{1995BaltA...4..245P}

Provencal, J.~L., Montgomery, M.~H., Kanaan, A., et~al., 2012, {\it \apj}, {\bf 751}, 91 \BIBCODE{2012ApJ...751...91P}

Pyrzas, S., G\"ansicke, B.~T., Hermes, J.~J., et~al., 2015, {\it \mnras}, {\bf 447}, 691 \BIBCODE{2015MNRAS.447..691P}

Quirion, P.-O., Fontaine, G., \& Brassard, P., 2007, {\it \apjs}, {\bf 171}, 219 \BIBCODE{2007ApJS..171..219Q}

Richer, H.~B. \& Ulrych, T.~J., 1974, {\it \apj}, {\bf 192}, 719 \BIBCODE{1974ApJ...192..719R}

Robinson, E.~L. \& McGraw, J.~T., 1976, {\it \apjl}, {\bf 207}, L37 \BIBCODE{1976ApJ...207L..37R}

Robinson, E.~L., Stover, R.~J., Nather, R.~E., \& McGraw, J.~T., 1978, {\it \apj}, {\bf 220}, 614 \BIBCODE{1978ApJ...220..614R}

Romero, A.~D., C\'orsico, A.~H., Althaus, L.~G., et~al., 2012, {\it \mnras}, {\bf 420}, 1462 \BIBCODE{2012MNRAS.420.1462R}

Romero, A.~D., Kepler, S.~O., C\'orsico, A.~H., Althaus, L.~G., \& Fraga, L., 2013, {\it \apj}, {\bf 779}, 58 \BIBCODE{2013ApJ...779...58R}

Sayres, C., Subasavage, J.~P., Bergeron, P., et~al., 2012, {\it \aj}, {\bf 143}, 103 \BIBCODE{2012AJ....143..103S}

Silvotti, R., Pavlov, M., Fontaine, G., Marsh, T., \& Dhillon, V., 2006, {\it \memsai}, {\bf 77}, 486 \BIBCODE{2006MmSAI..77..486S}

Silvotti, R., Voss, B., Bruni, I., et~al., 2005, {\it \aap}, {\bf 443}, 195 \BIBCODE{2005A&A...443..195S}

Steinfadt, J.~D.~R., Bildsten, L., Ofek, E.~O., \& Kulkarni, S.~R., 2008, {\it \pasp}, {\bf 120}, 1103 \BIBCODE{2008PASP..120.1103S}

Stobie, R.~S., Chen, A., O'Donoghue, D., \& Kilkenny, D., 1993, {\it \mnras}, {\bf 263}, L13 \BIBCODE{1993MNRAS.263L..13S}

Stobie, R.~S., Kilkenny, D., Koen, C., \& O'Donoghue, D., 1997, {\it The Third Conference 
on Faint Blue Stars}, ed. A.~G.~D. Philip, J.~Liebert, R.~Saffer, \& D.~S. Hayes, p. 497 \BIBCODE{1997fbs..conf..497S}

Stobie, R.~S., O'Donoghue, D., Ashley, R., et~al., 1995, {\it \mnras}, {\bf 272}, L21 \BIBCODE{1995MNRAS.272L..21S}

Stover, R.~J., Nather, R.~E., Robinson, E.~L., Hesser, J.~E., \& Lasker, B.~M., 1980, {\it \apj}, {\bf 240}, 865 \BIBCODE{1980ApJ...240..865S}

Stover, R.~J., Robinson, E.~L., \& Nather, R.~E., 1977, {\it \pasp}, {\bf 89}, 912 \BIBCODE{1977PASP...89..912S}

Su, J., Li, Y., \& Fu, J.-N., 2014a, {\it \na}, {\bf 33}, 52 \BIBCODE{2014NewA...33...52S}

Su, J., Li, Y., Fu, J.-N., \& Li, C., 2014b, {\it \mnras}, {\bf 437}, 2566 \BIBCODE{2014MNRAS.437.2566S}

Thompson, S.~E., Clemens, J.~C., \& Koester, D., 2005, {\it Astronomical Society of the Pacific 
Conference Series}, {\bf 334}, 14th European Workshop on White Dwarfs, ed. D.~Koester \& S.~Moehler, p. 471 \BIBCODE{2005ASPC..334..471T}

Thompson, S.~E., Clemens, J.~C., van Kerkwijk, M.~H., O'Brien, M.~S., \& Koester, D., 2004, {\it \apj}, {\bf 610}, 1001 \BIBCODE{2004ApJ...610.1001T}

Thompson, S.~E., Proven\c cal, J.~L., Kanaan, A., et~al., 2009, {\it Journal of Physics Conference Series}, {\bf 172}, 012067 \BIBCODE{2009JPhCS.172a2067T}

Tremblay, P.-E., Bergeron, P., \& Gianninas, A., 2011, {\it \apj}, {\bf 730}, 128 \BIBCODE{2011ApJ...730..128T}

Tremblay, P.-E., Ludwig, H.-G., Steffen, M., \& Freytag, B., 2013, {\it \aap}, {\bf 559}, A104 \BIBCODE{2013A&A...559A.104T}

Vauclair, G., Belmonte, J.~A., Pfeiffer, B., et~al., 1992, {\it \aap}, {\bf 264}, 547 \BIBCODE{1992A&A...264..547V}

Vauclair, G. \& Bonazzola, S., 1981, {\it \apj}, {\bf 246}, 947 \BIBCODE{1981ApJ...246..947V}

Vauclair, G., Dolez, N., \& Chevreton, M., 1981, {\it \aap}, {\bf 103}, L17 \BIBCODE{1981A&A...103L..17V}

Vauclair, G., Dolez, N., \& Chevreton, M., 1987, {\it \aap}, {\bf 175}, L13 \BIBCODE{1987A&A...175L..13V}

Vauclair, G., Dolez, N., Fu, J.~N., \& Chevreton, M., 1997, {\it \aap}, {\bf 322}, 155 \BIBCODE{1997A&A...322..155V}

Vauclair, G., Dolez, N., Fu, J.-N., et~al., 2000, {\it Baltic Astronomy}, {\bf 9}, 133 \BIBCODE{2000BaltA...9..133V}

Vauclair, G., Goupil, M.~J., Baglin, A., Auvergne, M., \& Chevreton, M., 1989, {\it \aap}, {\bf 215}, L17 \BIBCODE{1989A&A...215L..17V}

Voss, B., Koester, D., Ostensen, R., et~al., 2006, {\it \aap}, {\bf 450}, 1061 \BIBCODE{2006A&A...450.1061V}

Voss, B., Koester, D., Ostensen, R., et~al., 2007, {\it Astronomical Society of the Pacific 
Conference Series}, {\bf 372}, 15th European Workshop on White Dwarfs, ed. R.~Napiwotzki \& M.~R. Burleigh, p. 583 \BIBCODE{2007ASPC..372..583V}

Wenger, M., Ochsenbein, F., Egret, D., et~al., 2000, {\it \aaps}, {\bf 143}, 9 \BIBCODE{2000A&AS..143....9W}

Winget, D.~E., Nather, R.~E., Clemens, J.~C., et~al., 1990, {\it \apj}, {\bf 357}, 630 \BIBCODE{1990ApJ...357..630W}

Yeates, C.~M., Clemens, J.~C., Thompson, S.~E., \& Mullally, F., 2005, {\it \apj}, {\bf 635}, 1239 \BIBCODE{2005ApJ...635.1239Y}

\endreferences

\end{document}